\documentclass[10pt, aip, jcp, preprint, english, onecolumn, notitlepage]{revtex4-1}
\usepackage{graphicx}
\usepackage{graphics}
\usepackage{amsmath}
\usepackage{amssymb}
\usepackage{array}
\usepackage{dcolumn}
\usepackage{braket}
\usepackage[inline,draft,silent,nomargin]{fixme}
\usepackage{color}
\usepackage{bm} 
\usepackage{multirow}

\definecolor{blue1}{RGB}{0,0,128}
\usepackage[pdftitle={CC2 oscillator strengths within the local framework for calculating excitation energies (LoFEx)},
            pdfauthor={Pablo Baudin},
            pdfsubject={Report},
            colorlinks, 
            linkcolor=blue1,
            citecolor=blue1, 
            filecolor=blue1,
            urlcolor=blue1]{hyperref}
\usepackage{cleveref}
\crefname{equation}{eq.}{eqs.}%
\crefname{table}{table}{tables}%
\crefname{figure}{fig.}{figs.}%
\crefname{chapter}{chapter}{chapters}%
\crefname{section}{section}{sections}%
\crefname{appendix}{appendix}{appendices}%
\crefname{enumi}{point}{points}%

\usepackage{newfloat,algcompatible}
\usepackage[size=small]{caption}
\usepackage{etoolbox}

\AtBeginEnvironment{algorithm}{\noindent\hrulefill\par\nobreak\vskip-5pt}
\DeclareFloatingEnvironment[
      fileext=loa,
      listname=List of Algorithms,
      name=ALG.,
      placement=tbhp,
]{algorithm}
\DeclareCaptionFormat{algorithms}{\vskip-15pt\hrulefill\par#1#2#3\vskip-6pt\hrulefill}
\graphicspath{{./img/}}

\newcommand\ie{{\it i.e.}}
\newcommand\eg{{\it e.g.}}

\newcommand{\etal}{{\it et al.}}
\newcommand{\lofex}{LoFEx}
\newcommand{\Aeff}{\mathbf{A}^\text{eff}}
\newcommand{\den}{\epsilon_i - \epsilon_a + \epsilon_j - \epsilon_b}

\newcommand{\bibrad}[4]{\Bra{\widetilde{^{#1 #3}_{#2 #4}} }}

\newcommand{\up}[1]{\textsuperscript{#1}}
\newcommand{\cart}{\mathfrak{j}}

\fxsetup{theme=color}
\definecolor{fxnote}{RGB}{200,0,0}
\definecolor{fxtarget}{RGB}{30,30,255}

\begin{document}
\title{CC2 oscillator strengths within the local framework for calculating excitation energies (\lofex{})}
\date{\today}
\author{Pablo Baudin}
\email[]{pablo.baudin@chem.au.dk}
\author{Thomas Kj{\ae}rgaard}
\author{Kasper Kristensen}
\affiliation{qLEAP Center for Theoretical Chemistry, Department of Chemistry, Aarhus University, Langelandsgade 140, DK-8000 Aarhus C, Denmark}

\begin{abstract}
In a recent work [Baudin and Kristensen, \emph{J. Chem. Phys.} \textbf{144}, 224106 (2016)], 
we introduced a local framework for calculating excitation energies (\lofex{}), based on
second-order approximated coupled cluster (CC2) linear-response theory.
\lofex{} is a black-box method in which a reduced excitation orbital space (XOS) is optimized
to provide coupled cluster (CC) excitation energies at a reduced computational cost. 
In this article, we present an extension of the \lofex{} algorithm to the calculation
of CC2 oscillator strengths. Two different strategies are suggested, in which the size of the XOS  
is determined based on the excitation energy or the oscillator strength of the targeted transitions. 
The two strategies are applied to a set of medium-sized organic molecules in order to assess both the
accuracy and the computational cost of the methods. 
The results show that CC2 excitation energies and oscillator strengths can be calculated at
a reduced computational cost, provided that the targeted transitions are local compared
to the size of the molecule. 
To illustrate the potential of \lofex{} for large molecules, both strategies have been 
successfully applied to the lowest transition of the bivalirudin molecule ($4255$ basis functions)
and compared with time-dependent density functional theory.
\end{abstract}

\maketitle

\section{Introduction}

High-accuracy calculations of electronic absorption spectra can be performed using coupled
cluster (CC) response theory\cite{Koch1995,Christiansen1996,Christiansen1998a} 
via the computation of excitation energies and oscillator strengths. 
CC theory is well established as the method of choice for describing the 
electronic structure of molecules with a ground-state dominated by a single electronic 
configuration. However, the high-accuracy of CC models comes with a high computational cost and for that 
reason standard CC calculations of excitation energies and oscillator strengths have been 
limited to rather small molecules. Less reliable computational models like time-dependent 
density functional theory (TDDFT) are thus extensively used for the simulation of 
electronic spectra of medium-sized and large molecules.\cite{Gonzalez2012}
We note that the equation-of-motion (EOM) CC formalism is closely related
to CC response theory and is often used in the same context.\cite{Stanton1993,Watts1995a,Gwaltney1996} 
While EOM and response techniques are identical for the calculation of CC excitation energies,
we have chosen to consider CC response theory in this work since it results in size-intensive 
transition moments, in contrast to EOM-CC theory.\cite{Koch1994}

The computational scaling of CC methods with the system size is associated with the 
usage of canonical Hartree-Fock (HF) orbitals which are generally delocalized in space, while
CC theory describes local phenomena (electron correlation effects).\cite{Saeboe1993} In the last decades,
a lot of efforts have been dedicated to the design of low-scaling CC models, primarily
for the computation of ground-state energies.\cite{Saeboe1993,Werner2011,Yang2012,Riplinger2013a,Eriksen2015,Werner2015,Riplinger2016,Li2016,Nagy2016,Friedrich2012} 
More recently, several groups turned their attention to the calculation of excitation energies
and molecular properties using local approximations. The combination of local occupied orbitals with
non-orthogonal virtual orbitals (\eg{} projected atomic orbitals (PAOs) or pair natural orbitals (PNOs)) 
is widely used to reduce the total number of wave function parameters and it has been applied to 
the calculation of excitation energies,\cite{Crawford2002,Korona2003,Kats2006,Kats2009,Kats2010,Helmich2013,Helmich2014,Dutta2016} transition strengths,\cite{Kats2007,Kats2010} and other molecular properties.\cite{Kats2007,Crawford2010,Kats2010,Russ2008,McAlexander2016} 
The incremental scheme in which the quantities of interest are
expanded in a many-body series has also been applied to the calculation of CC excitation energies\cite{Mata2011} 
and dipole polarizabilities.\cite{Friedrich2015} Another recent development is
the multilevel CC theory in which different CC models are used to treat different parts of the system.\cite{Myhre2013,Myhre2014,Myhre2016}
In this context, we can also mention the reduced virtual space \cite{Send2011} and ONIOM strategies.\cite{Caricato2010,Caricato2011a}

In a recent publication,\cite{Baudin2016a} we have introduced a new strategy for the calculation of CC excitation energies
at a reduced computational cost, in which we focused on the second-order approximated CC singles and doubles 
(CC2) model.
In our local framework for calculating excitation energies (\lofex{}), the locality of correlation 
effects is used to generate a state-specific 
mixed orbital space composed of the dominant pair of natural transition orbitals (NTOs),
obtained from time-dependent Hartree-Fock (TDHF) theory, and localized molecular orbitals (LMOs).
This mixed orbital space is well adapted to describe the targeted electronic transition 
and can be significantly reduced (by discarding a subset of least relevant LMOs in a black-box manner) 
without affecting the accuracy of the calculated excitation energy. In this way, important computational 
savings are possible for local transitions in large molecular systems.

In \Cref{sec:ricc2}, we briefly summarize how excitation energies and oscillator strengths can be computed at the CC2 level 
of theory. The \lofex{} algorithm for excitation energies is then summarized in \Cref{sec:lofex}, in which we also suggest two different strategies 
for computing oscillator strengths within \lofex{}. 
In \Cref{sec:results}, these strategies are compared when applied to the lowest electronic 
transitions of a set of medium-sized organic molecules. We also present results for a large
molecule (bivalirudin) and compare the accuracy and computational efforts of \lofex{} with 
TDDFT/CAM-B3LYP calculations.

\section{The RI-CC2 model for oscillator strengths}
\label{sec:ricc2}

The CC2 model was introduced by Christiansen \etal{}\cite{Christiansen1995} as an
intermediate model between the CCS and CCSD models in the CC hierarchy for the calculation
of frequency-dependent properties. CC2 is therefore the first model of the CC hierarchy to 
include correlation effects and thus constitutes an appropriate starting point for \lofex{}.
In this section, we summarize how CC2 excitation energies and oscillator strengths can be 
obtained from response theory.

The CC2 ground-state amplitudes are obtained as solution of the following non-linear equations,\cite{Christiansen1995}
\begin{eqnarray}
   \label{eq:amp1}
   \Omega_{\mu_1} &=& \bra{\mu_1} \hat{H} + [\hat{H}, T_2] \ket{\text{HF}} = 0, \\
   \Omega_{\mu_2} &=& \bra{\mu_2} \hat{H} + [F, T_2] \ket{\text{HF}} = 0, \label{eq:amp2}
\end{eqnarray}
where $\{ \ket{\text{HF}},\ket{\mu_1}, \ket{\mu_2}\}$ denote the Hartree-Fock (HF)
ground-state and the set of singles and doubles excitation manifolds.
$F$ is the Fock operator and $\hat{H}$ is a similarity (T$_1$)-transformed Hamiltonian,
\begin{equation}
   \label{eq:t1tr}
   \hat{H} = \exp(-T_1) H \exp(T_1),
\end{equation}
where $T_i=\sum_{\mu_i} t_{\mu_i} \tau_{\mu_i}$ is a cluster operator, $t_{\mu_i}$
is a cluster amplitude, $\tau_{\mu_i}$ is an excitation operator, and $i$
denotes the excitation level.
The $T_1$-transformation of the Hamiltonian in \Cref{eq:t1tr} can be transferred to the second-quantization
elementary operators, which effectively corresponds to a modification of the molecular
orbital (MO) transformation matrices $\mathbf{C}$ with the singles amplitudes,\cite{Koch1994a,mest}
\begin{equation}
   \begin{array}{rcl}
      X_{\alpha i} &=& C_{\alpha i} \\ 
      Y_{\alpha i} &=& C_{\alpha i} + \sum_a C_{\alpha a} t^a_i 
   \end{array}\qquad
   \begin{array}{rcl}
      X_{\alpha a} &=& C_{\alpha a} - \sum_i C_{\alpha i} t^a_i \\
      Y_{\alpha a} &=& C_{\alpha a}
   \end{array}
\end{equation}
A two-electron $T_1$-transformed integral in the Mulliken notation can now be expressed as,
\begin{align}
(pq\hat{|}rs) =& \sum_{\alpha \beta \gamma \delta} X_{\alpha p} Y_{\beta q} X_{\gamma r} Y_{\delta s} (\alpha \beta | \gamma \delta),
\end{align}
where we have used the following convention to denote orbitals:
\begin{itemize}
	\item Atomic orbitals (AOs): $\alpha, \beta, \gamma \dots$
	\item MOs of unspecified occupancy: $p, q, r \dots$
	\item Occupied MOs: $i, j, k \dots$
	\item Virtual MOs: $a, b, c \dots$
\end{itemize}
Since only closed-shell molecules are targeted in this work, all MOs are considered spin-free.

In the CC2 model, the doubles amplitudes are only correct
through first-order in the fluctuation potential ($\Phi = H - F$). This approximation
leads to a closed-form of the doubles amplitudes,
\begin{align}
	t^{ab}_{ij} = \frac{(ai\hat{|}bj)}{\den{}},
\end{align}
where $\epsilon_p$ denotes the orbital energy associated with orbital $p$.
The CC2 equations can then be formulated in a CCS-like manner in which the doubles
amplitudes are calculated on-the-fly. In order to take full advantage of this formulation
and avoid the storage of any four-index quantity (amplitudes or integrals), H{\"a}ttig and Weigend
used the resolution-of-the-identity (RI) approximation for the two-electron integrals\cite{Whitten1973,Feyereisen1993a} both 
in the optimization of the CC2 ground-state and excitation amplitudes.\cite{Hattig2000}
This strategy was later generalized to the calculation of transition strengths and 
excited-state first-order properties.\cite{Hattig2002}

In CC response theory, excitation energies and transition strengths 
from the ground-state ($0$) to an excited-state ($m$) are obtained from the poles and 
residues of the linear-response function, respectively.\cite{Christiansen1998} The poles of the
CC linear-response function correspond to the eigenvalues of the non-symmetric 
Jacobian matrix,
\begin{equation}
A_{\mu_i \nu_j}=\partial \Omega_{\mu_i}/\partial t_{\nu_j},
\end{equation}
while electric dipole transition strengths are given by,
\begin{align}
   S_{0m}^{V^\cart V^\cart} &= T_{0m}^{V^\cart} T_{m0}^{V^\cart}, \label{eq:trans} \\
   T_{0m}^{V^\cart} &= \sum_{pq} [ D_{pq}^{\eta} (\mathbf{R}) + D_{pq}^{\xi} (\bar{\mathbf{M}}) ] \hat{V}^\cart_{pq}, \label{eq:rmom} \\
   T_{m0}^{V^\cart} &= \sum_{pq} D_{pq}^{\xi} (\mathbf{L})  \hat{V}^\cart_{pq}, \label{eq:lmom} 
\end{align}
where $\hat{V}^\cart_{pq}$ is a Cartesian component ($\cart=x,y,z$) of the T$_1$-transformed electric dipole integrals in the length gauge
and $D_{pq}^{\eta}$ and $D_{pq}^{\xi}$ are one-electron density matrices (see \Cref{ap:cc2eq}). 
$\mathbf{R}$ and $\mathbf{L}$ are the right and left Jacobian eigenvectors following
the normalization condition $\mathbf{L} \mathbf{R} =1$ and $\bar{\mathbf{M}}$ are the
transition moment Lagrangian multipliers. In addition, the ground-state Lagrangian 
multipliers $\bar{\mathbf{t}}$ are required for the calculation of the one-electron
density matrices. As for the CC2 ground-state amplitudes, the CC2 excitation amplitudes
and the Lagrange multipliers can be obtained without storing any four-index quantity 
by considering an effective Jacobian matrix,\cite{Hattig2000,Hattig2002}
\begin{equation}
   A^{\text{eff}}_{\mu_1 \nu_1} (\omega)  =  A_{\mu_1 \nu_1} - \sum_{\gamma_2} 
      \frac{ A_{\mu_1 \gamma_2} A_{\gamma_2 \nu_1} }{\epsilon_{\gamma_2} - \omega},
   \label{eq:jac}
\end{equation}
where $\omega$ is an excitation energy and $\epsilon_{aibj} = \epsilon_a - \epsilon_i + \epsilon_b - \epsilon_j$. 
Using the effective Jacobian, the response equations to be solved become,
\begin{align}
&  \Aeff{}(\omega) \mathbf{R}_1 = \omega \mathbf{R}_1, \label{eq:xright} \\
&  \mathbf{L}_1 \Aeff{}(\omega) = \mathbf{L}_1 \omega, \label{eq:xleft} \\
&  \bar{\mathbf{t}}_1 \Aeff{}(0) = -\bm{\eta}^\text{eff}_1, \label{eq:tbar} \\
&  \bar{\mathbf{M}}_1 (\Aeff{}(-\omega) + \omega \mathbf{1}) = -\bar{\mathbf{m}}^\text{eff}_1, \label{eq:mbar} 
\end{align}
where the subscript $1$ denotes the singles part of a vector, and $\bm{\eta}^\text{eff}_1$ and
$\bar{\mathbf{m}}^\text{eff}_1$ are the effective right-hand-sides of the linear equations
for the ground-state and transition moments Lagrange multipliers, respectively.
Once \Cref{eq:xright,eq:xleft,eq:tbar,eq:mbar} have been solved, the one-electron density matrices
$D_{pq}^{\eta}$ and $D_{pq}^{\xi}$ can be calculated and contracted with $\hat{V}^\cart_{pq}$
to get the right and left transition dipole moments as well as transition strengths.
All doubles quantities can be constructed on-the-fly whenever needed using the RI approximation
for the two-electron integrals.
In \Cref{ap:cc2eq} we collect all the working equations required to calculate CC2 
transition moments in a canonical MO basis. The equations are given here for completeness but
should be equivalent to the ones in Ref. \onlinecite{Hattig2002}. (A few typos were present
in the original paper and are corrected in \Cref{ap:cc2eq}).

When studying electronic transitions, one often consider oscillator strengths instead of
the transition strengths given by \Cref{eq:trans}. Oscillator strengths in the length gauge 
are straightforwardly obtained as,
\begin{align}
   f_{0m} = \frac{2}{3} \omega_m \sum_{\cart=x,y,z} S_{0m}^{V^\cart V^\cart},
\end{align}
where $\omega_m$ is the excitation energy for a transition from the ground-state to the $m$-th excited-state.
The calculation of excitation energies and oscillator strengths at the CC2 level has been 
implemented in a local version of the \textsc{LSDalton} program\cite{dalton,lsdalton} following 
the strategy presented in Refs. \onlinecite{Hattig2000} and \onlinecite{Hattig2002}.

\section{Excitation energies and oscillator strengths within \lofex{}}
\label{sec:lofex}

In a previous publication\cite{Baudin2016a} we have introduced the \lofex{} algorithm 
as a framework to calculate CC2 excitation energies of large molecules.
In this section, we summarize the \lofex{} procedure and extend it to the computation of 
CC2 oscillator strengths. 

\subsection{Excitation energies}

\begin{figure}
	\includegraphics[width=0.4\textwidth]{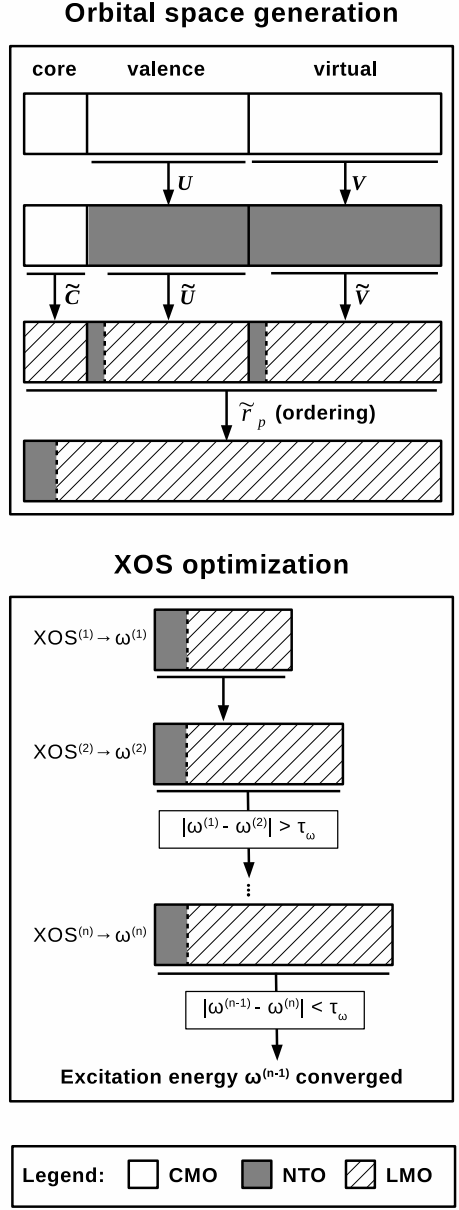}
	\caption{Schematic representation of the original \lofex{} procedure. 
      $\mathbf{U}$ and  $\mathbf{V}$ represent the transformation matrices from canonical molecular 
      orbitals (CMOs in white) to natural transition orbitals (NTOs in grey) for the valence and virtual spaces, respectively. 
      $\mathbf{\widetilde{C}}$, $\mathbf{\widetilde{U}}$, and $\mathbf{\widetilde{V}}$ represent the 
      transformation matrices to local molecular orbitals (LMOs in stripes) for core, valence, and 
      virtual orbitals, respectively, excluding the dominant pair of NTOs.   
      $\tilde{r}_p$ is the effective distance measure given by \Cref{eq:list}, $\omega^{(i)}$ is
      the excitation energy corresponding to the $i$-th excitation orbital space (XOS$^{(i)}$)
        and $\tau_\omega$ is the \lofex{} excitation energy threshold.
   }
   \label{fig:lofex_summary}
\end{figure}

In \lofex{} a transition-specific orbital space is constructed based on the solutions of the 
TDHF problem for the whole molecule and starting from HF canonical molecular orbitals (CMOs).
First, NTOs are obtained by performing a singular-value-decomposition (SVD) of the
TDHF transition density matrix, $\mathbf{\widetilde{b}}$, for each transition of interest,\cite{Luzanov1976,Etienne2015,Baudin2016a}
\begin{align}
   \label{eq:trU}
   \mathbf{\widetilde{b}}^{\dagger} \mathbf{\widetilde{b}} \mathbf{u}_k &= \lambda_k \mathbf{u}_k, \quad k=1,2,\cdots,N_o, \\
   \label{eq:trV}
   \mathbf{\widetilde{b}} \mathbf{\widetilde{b}}^{\dagger} \mathbf{v}_k &= \lambda_k' \mathbf{v}_k, \quad k=1,2,\cdots,N_v, 
\end{align}
which leads to the transformation matrices from CMOs to NTOs for the occupied and virtual spaces, respectively,
\begin{align}
   \mathbf{U} &= (\mathbf{u}_1, \mathbf{u}_2, \cdots, \mathbf{u}_{N_o}), \\
   \mathbf{V} &= (\mathbf{v}_1, \mathbf{v}_2, \cdots, \mathbf{v}_{N_v}),
\end{align}
Where $N_o$ ($N_v$) is the number of occupied valence (virtual) orbitals.
Assuming $N_o \le N_v$, it follows that $\lambda_k \equiv \lambda_k' $ for $k=1,2,\cdots,N_o$, while $\lambda_k' = 0$
for $k=N_o+1,\cdots,N_v$. 
The relevance of a given pair of NTOs ($k$) in the electronic 
transition associated with the density matrix $\mathbf{\widetilde{b}}$ can be evaluated through its 
singular value $\sqrt{\lambda_k}$.\cite{Luzanov1976,Martin2003}
For singles-dominated transitions, one pair of NTOs (with singular value close to one) dominates
the transition, while the other NTOs are far less important to describe the process and thus have
much smaller singular values. In \lofex{}
we therefore keep the dominant pair of NTOs intact, while the remaining orbitals are localized using
the square of the second central moment of the orbitals as a localization function.\cite{Jansik2011,Hoeyvik2012b} This 
procedure (summarized in the upper part of \Cref{fig:lofex_summary}) results
in a mixed orbital space composed of orthogonal NTOs and localized molecular orbitals (LMOs) that
is adapted to the description of a specific electronic transition. Core orbitals are not considered
in the generation of NTOs and are localized independently to avoid mixing between core and valence
spaces.

In this mixed orbital space, the dominant pair of NTOs is expected to describe the main character of the
targeted electronic transition, while the LMOs enable an efficient description of correlation effects.
In order to reduce the computational cost of the CC calculation, a subspace of the mixed NTO/LMO 
space is then constructed by considering the most relevant orbitals based on an effective
distance $\tilde{r}_p$ given by,
\begin{equation}
   \tilde{r}_p = \min_A \left( \frac{r_{Ap}}{Q^\text{NTO,o}_A}, \frac{r_{Ap}}{Q^\text{NTO,v}_A}\right),
   \label{eq:list}
\end{equation}
where index $A$ denotes atomic centers, $r_{Ap}$ corresponds to the distance between the
center of charge of a local orbital $p$ and atomic center $A$, and $Q^\text{NTO,o}_A$
and $Q^\text{NTO,v}_A$ are the L{\"o}wdin atomic charges of the occupied and virtual NTOs on
center $A$, respectively. The resulting reduced space is denoted the excitation
orbital space (XOS). The inactive Fock matrix can then be diagonalized in the XOS to obtain a set of pseudo-canonical
orbitals. CC excitation energies (and eventually oscillator strengths) can then be calculated
in the XOS using standard canonical implementations, as described in \Cref{sec:ricc2} and \Cref{ap:cc2eq} for the CC2 model.

In order to preserve the black-box feature of CC theory, the XOS is optimized as
depicted in the lower part of \Cref{fig:lofex_summary}, \ie{}, a first guess for the 
XOS (XOS$^{(1)}$) is built and the CC problems are solved in that space to provide the
excitation energy $\omega^{(1)}$, the XOS is then extended based on the list defined by \Cref{eq:list} until
the difference between the last two excitation energies is smaller than the \lofex{} excitation energy threshold $\tau_\omega$,
($|\omega^{(n-1)} - \omega^{(n)}| < \tau_\omega$). 
We have shown in Ref. \onlinecite{Baudin2016a} (where $\tau_\omega$ was denoted $\tau_\text{XOS}$), that this procedure can result in
significant speed-ups compared to standard CC2 implementations without loss of
accuracy. 

\subsection{Oscillator strengths}

For the calculation of oscillator strengths with \lofex{}, we consider the following strategies:
\begin{enumerate}
   \item The XOS is optimized solely based on the excitation energy (as 
      described in \Cref{fig:lofex_summary} and Ref. \onlinecite{Baudin2016a}) and 
      the oscillator strength is only calculated once in the optimized XOS (XOS$^{(n-1)}$).
      \label{it:standard}
   \item Both excitation energies and oscillator strengths are calculated in
      each \lofex{} iteration and only the oscillator strength is checked for
      convergence. In other words, the XOS is considered converged when,
      $| f^{(n)} - f^{(n-1)} | < \tau_f$, where $\tau_f$ is the \lofex{} oscillator strength threshold.
      \label{it:spectrum}
\end{enumerate}
Note that in the XOS optimization, the last step (step $n$) is necessary to check that step
$n-1$ was already converged. The calculation of oscillator strengths in \cref{it:standard}
is therefore done in the penultimate XOS to ensure minimal computational efforts.

In the following section, we will refer to \cref{it:standard} as the standard-\lofex{} strategy, while \cref{it:spectrum} 
is denoted the spectrum strategy. Indeed, in \cref{it:spectrum} the oscillator strength threshold
$\tau_f$ has a different purpose than the excitation energy threshold $\tau_\omega$. Checking only the oscillator strength
for convergence is expected to provide a balanced description of the transitions
in the sense that transitions with large oscillator strengths should be well described, while 
weak transitions (with $f \simeq 0$) are expected to converge in minimal XOSs and
lead to less accurate excitation energies, while using less computational resources.  
The standard-\lofex{} strategy is thus preferred if accurate excitation energies are 
requested for all transitions, while the spectrum strategy is more appropriate if one
is only interested in transitions with a significant oscillator strengths.

\section{Results}
\label{sec:results}

In this section we present numerical results for excitation energies and oscillator strengths
using the standard- and spectrum-\lofex{} strategies introduced in \Cref{sec:lofex}. 
For that purpose, we consider the following set of medium-sized organic molecules,
\begin{itemize}
   \item caprylic acid,
   \item lauric acid,
   \item palmitic acid,
   \item 15-oxopentadecanoic acid (15-OPDA),
   \item prostacyclin,
   \item an $\alpha$-helix composed of 8 glycine residues ($\alpha$-Gly$_8$),
   \item leupeptin,
   \item latanoprost,
   \item met-enkephalin, and
   \item 11-cis-retinal.
\end{itemize}
The molecular geometry for 11-cis-retinal was obtained from Ref. \onlinecite{Dutta2016},
while for the other systems, the Cartesian coordinates as well as details regarding the optimization of the structures are
available in Ref. \onlinecite{Baudin2016a} and its supporting information. All the 
calculations presented in this section have been performed with a local version of 
the LSDalton program,\cite{lsdalton,dalton} using the correlation consistent
aug-cc-pVDZ' basis set\cite{dunning1989,dunning1992} with the corresponding auxiliary
basis, aug-cc-pVDZ-RI' for the RI approximation.\cite{Hattig2002a} The prime in the
basis set notation indicates that diffuse functions have been removed on the hydrogen atoms.

The parameters used in the following investigation have been set to the same default values
as in Ref. \onlinecite{Baudin2016a}, \ie{}, the \lofex{} excitation energy threshold was set to 
$\tau_\omega = 0.02$ eV and the number of orbitals added to the XOS in each \lofex{}
iteration corresponds to ten times the average number of orbitals per atom.
For the spectrum-\lofex{} strategy we have chosen $\tau_f = 0.001$.

\subsection{Calculation of oscillator strengths within \lofex{}}
\label{sec:res1}

\begin{table*}
   \caption{
      Comparison of standard-\lofex{} ($\tau_\omega = 0.02$ eV) and conventional CC2 excitation 
      energies and oscillator strengths.
      The \lofex{} excitation energies are given in eV for the largest XOS (step $n$) and for the 
      converged XOS (step $n-1$), while oscillator strengths are only reported for the converged XOS.
      Absolute errors are given for both excitation energies and oscillator strengths.
      Finally, the number of iterations in the XOS optimization ($n$) as well as speed-ups
      of \lofex{} compared to conventional CC2 algorithms are also reported.
      \label{tab:standard}
   }
   \begin{ruledtabular}
   \begin{tabular}{lccccccccc}
      System&State&No. iter. $(n)$&$\omega^{(n)}$&$\delta \omega^{(n)}$&$\omega^{(n-1)}$&$\delta \omega^{(n-1)}$&$f^{(n-1)}$&$\delta f^{(n-1)}$&Speed{-}up\\
      \hline
      Caprylic acid       &S$_1$&2&6.06&0.00&6.07&0.01&0.000&0.000&\multirow{2}{*}{0.72}\\
                          &S$_2$&3&6.83&0.00&6.83&0.00&0.066&0.003& \\
      Lauric acid         &S$_1$&3&6.05&0.00&6.06&0.00&0.000&0.000&\multirow{2}{*}{1.29}\\
                          &S$_2$&3&6.81&0.00&6.82&0.01&0.065&0.005& \\
      Palmitic acid       &S$_1$&3&6.06&0.00&6.06&0.00&0.000&0.000&\multirow{2}{*}{4.07}\\
                          &S$_2$&3&6.81&0.01&6.83&0.03&0.065&0.005& \\
      15{-}OPDA           &S$_1$&2&4.44&0.00&4.45&0.01&0.000&0.000&\multirow{3}{*}{1.52}\\
                          &S$_2$&2&6.06&0.00&6.08&0.02&0.000&0.000& \\
                          &S$_3$&5&6.19&0.00&6.19&0.00&0.040&0.001& \\
      Prostacyclin        &S$_1$&5&4.98&0.00&4.99&0.01&0.005&0.000&1.16\\
      $\alpha$-gly$_8$    &S$_1$&4&5.43&0.00&5.43&0.00&0.001&0.000&\multirow{2}{*}{2.46}\\
                          &S$_2$&4&5.73&0.00&5.73&0.01&0.003&0.001& \\
      Leupeptin           &S$_1$&3&4.27&0.01&4.28&0.01&0.001&0.000&3.37\\
      Latanoprost         &S$_1$&3&5.08&0.00&5.08&0.01&0.001&0.000&16.8\\
      Met{-}enkephalin    &S$_1$&3&4.78&0.00&4.79&0.01&0.024&0.002&34.0\\
      11-cis-retinal      &S$_1$&5&2.14&0.00&2.14\footnote{The full molecule was included in step $n$ which was not yet converged, 
   so in this case, $\omega^{(n-1)}$ and $f^{(n-1)}$ are effectively calculated in XOS$^{(n)}$.}
      &0.00\footnotemark[1]&1.384\footnotemark[1]&0.000\footnotemark[1]&0.61\\
   \end{tabular}
   \end{ruledtabular}
\end{table*}

In \Cref{tab:standard} we report the \lofex{} excitation energies and oscillator strengths for the
lowest electronic transitions of the molecules presented above when using the standard-\lofex{} strategy.
Absolute errors in the excitation energies and the oscillator strengths as well as speed-ups
compared to conventional CC2 implementations are also reported.
Since in the standard-\lofex{} strategy, the oscillator strength is only calculated in the converged 
(penultimate) XOS, we report excitation energies corresponding to both the expanded (step $n$) 
and converged (step $n-1$) XOSs. We note that, as demonstrated in Ref. \onlinecite{Baudin2016a},
the excitation energies in the expanded steps are ``overconverged'' (all errors are well below $0.02$ eV),
while in the penultimate steps the errors in the excitation energies are of the order of the
\lofex{} excitation energy threshold ($0.02$ eV). For the oscillator strengths, the absolute errors are equal or
below $0.005$ and are strongly correlated with the intensity of the transitions (larger oscillator
strengths correspond to larger errors), except for 11-cis-retinal which include the complete 
orbital space.

\begin{figure}
	\includegraphics[width=0.5\textwidth]{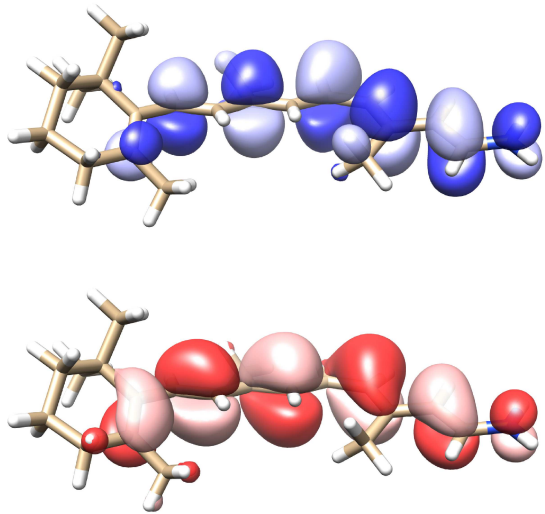}
	\caption{Stick representation of the 11-cis-retinal molecule. Natural transition orbitals 
      for the lowest transition are represented with a contour value of $0.02$ a.u. 
      (bottom: occupied NTO, top: virtual NTO).\cite{chimera}
   }
   \label{fig:retinal}
\end{figure}

Regarding the speed-ups of the standard-\lofex{} algorithm compared to a conventional (multi-state) CC2 implementation, it is
found that the state-specific approach of \lofex{} remains advantageous in most cases, even for the computation
of several transitions. This is of course strongly dependent on the character of the transitions
and on the size of the molecule, \eg{} in the case of 15-OPDA, the two lowest transitions are rather
local and converge in only two \lofex{} iterations but the third transition has a more delocalized
character\cite{Baudin2016a} and requires almost the complete orbital space to be included in the XOS which limits significantly
the obtained speed-up ($1.52$). Another less favorable case for \lofex{} is the lowest transition
of 11-cis-retinal. Both the excitation energy and the oscillator strength are perfectly recovered
by \lofex{}. However, the complete orbital space is required in order to determine the excitation energy
to the desired precision, and the oscillator strength thus also has to be calculated in
the complete orbital space, which results in a ``speed-up'' of $0.61$. This behaviour can be understood
by looking at the dominant pair of NTOs in \Cref{fig:retinal}, which shows that the transition is basically
affecting the whole molecule, preventing any computational savings using \lofex{}. This should be put
in contrast with the performance of \lofex{} for the met-enkephalin molecule, where both the
excitation energy and the oscillator strength are well described with only 3 \lofex{} iterations,
resulting in a significant speed-up ($34$). 
It should be emphasized that the gain in terms of computational efforts for met-enkephalin is much
greater than the computational overhead observed for 11-cis-retinal.
These two examples demonstrate that \lofex{} is designed to ensure error control and accuracy of the results,
while computational savings are transition and system dependent.

\begin{table*}
   \caption{
      Comparison of spectrum-\lofex{} ($\tau_f = 0.001$) and conventional CC2 excitation 
      energies and oscillator strengths.
      The \lofex{} excitation energies and the corresponding absolute errors are given in eV.
      We also report the oscillator strengths and corresponding absolute errors as well as 
      the number of iterations used in the XOS optimization ($n$) and the speed-ups
      of \lofex{} compared to conventional CC2 algorithms.
      \label{tab:spectrum}
   }
   \begin{ruledtabular}
   \begin{tabular}{lccccccc}
      System&State&No. iter. $(n)$&$\omega^{(n)}$&$\delta \omega^{(n)}$&$f^{(n)}$&$\delta f^{(n)}$&Speed{-}up\\
      \hline
      Caprylic acid       &S$_1$ &1&6.07&0.01&0.000&0.000&\multirow{2}{*}{0.65}\\
                          &S$_2$ &3&6.82&0.00&0.069&0.000& \\
      Lauric acid         &S$_1$ &1&6.08&0.02&0.000&0.000&\multirow{2}{*}{0.79}\\
                          &S$_2$ &4&6.81&0.00&0.070&0.000& \\
      Palmitic acid       &S$_1$ &1&6.09&0.04&0.000&0.000&\multirow{2}{*}{0.88}\\
                          &S$_2$ &5&6.80&0.00&0.070&0.000& \\
      15{-}OPDA           &S$_1$ &1&4.45&0.01&0.000&0.000&\multirow{3}{*}{0.97}\\
                          &S$_2$ &1&6.08&0.02&0.000&0.000& \\
                          &S$_3$ &5&6.19&0.00&0.040&0.001& \\
      Prostacyclin        &S$_1$ &5&4.98&0.00&0.005&0.000&0.71\\
      $\alpha$-gly$_8$    &S$_1$ &2&5.46&0.03&0.001&0.000&\multirow{2}{*}{11.1}\\
                          &S$_2$ &2&5.76&0.04&0.004&0.002& \\
      Leupeptin           &S$_1$ &1&4.30&0.04&0.000&0.001&69.0\\
      Latanoprost         &S$_1$ &3&5.07&0.00&0.001&0.000&9.31\\
      Met{-}enkephalin    &S$_1$ &4&4.78&0.00&0.022&0.000&5.01\\
      11-cis-retinal      &S$_1$ &5&2.14&0.00&1.384&0.000&0.42\\
   \end{tabular}
   \end{ruledtabular}
\end{table*}

With the idea of producing electronic spectra of CC2 quality at a reduced computational cost, we now
turn our attention to the spectrum-\lofex{} strategy. In electronic spectra, it is important to 
provide a good description of the transitions with large oscillator strengths and, for that purpose,
the standard-\lofex{} strategy might be inappropriate since it converges the XOS based on the excitation energies and
not on the oscillator strengths. 
In \Cref{tab:spectrum} we report the \lofex{} excitation energies and oscillator strengths for the
lowest electronic transitions of the molecules presented above when the spectrum-\lofex{} strategy is used
with $\tau_f = 0.001$.
Absolute errors in the excitation energies and the oscillator strengths as well as speed-ups
compared to conventional CC2 implementations are also reported.
Since in the spectrum-\lofex{} strategy, the oscillator strengths and excitation energies are 
calculated in each \lofex{} iteration, we only report the values corresponding to the most 
accurate results, \ie{}, the ones from the expanded XOS (step $n$).
Note also that, while in the standard-\lofex{} procedure at least two steps are necessary to 
check the convergence of excitation energies ($n\ge2$), in the spectrum strategy we consider that the first
step can be directly converged if $f^{(1)} < \tau_f$.

From \Cref{tab:spectrum}, we see that for the strongest transitions (with $f > 0.01$) the errors
in both the excitation energies and the oscillator strengths are very satisfactory. As expected,
for weaker transitions, larger errors occur in the excitation energies (up to $0.04$ eV) which 
is related to the fact that only the oscillator strengths are used to converge the XOS.
For example, for the lowest transition of leupeptin ($f=0.001$ and $\delta \omega = 0.04$), 
the weak character of the transition leads to a converged XOS in the first iteration and a 
significant speed-up ($69.0$) is observed. 
However, the results in \Cref{tab:spectrum} also show that, even if some
computational time is saved on the weakest transitions, more time has to be dedicated to the 
stronger ones since larger XOSs are required to achieved the desired accuracy and since 
the oscillator strengths have to be calculated in each \lofex{} iteration.
As a consequence, less impressive speed-ups are observed for the spectrum-\lofex{} strategy (except for $\alpha$-gly$_8$ and leupeptin).
However, as for the standard-\lofex{} strategy in \Cref{tab:standard} we note that the potential speed-ups
are much larger than the additional overhead present in the less favorable cases.

Comparing \Cref{tab:standard,tab:spectrum} we note that for all the transitions with $f > 0.01$,
the number of required iterations with the spectrum strategy is always larger or equal to the
number of iterations used in the standard-\lofex{} strategy.
In accordance with Ref. \onlinecite{Kats2007}, this suggests that a fine-tuned description of
(strong) oscillator strengths requires larger orbital spaces than the excitation energy alone.
Finally, we note that both the accuracy and the computational savings are driven by the main \lofex{}
threshold ($\tau_f$ for the spectrum strategy) and that in practical applications of \lofex{},
$\tau_f$ could of course be increased to reduce the computational efforts at the expense of 
obtaining slightly less accurate oscillator strengths.

\subsection{Large-scale application: the bivalirudin molecule}

\begin{figure}
	\includegraphics[width=0.5\textwidth]{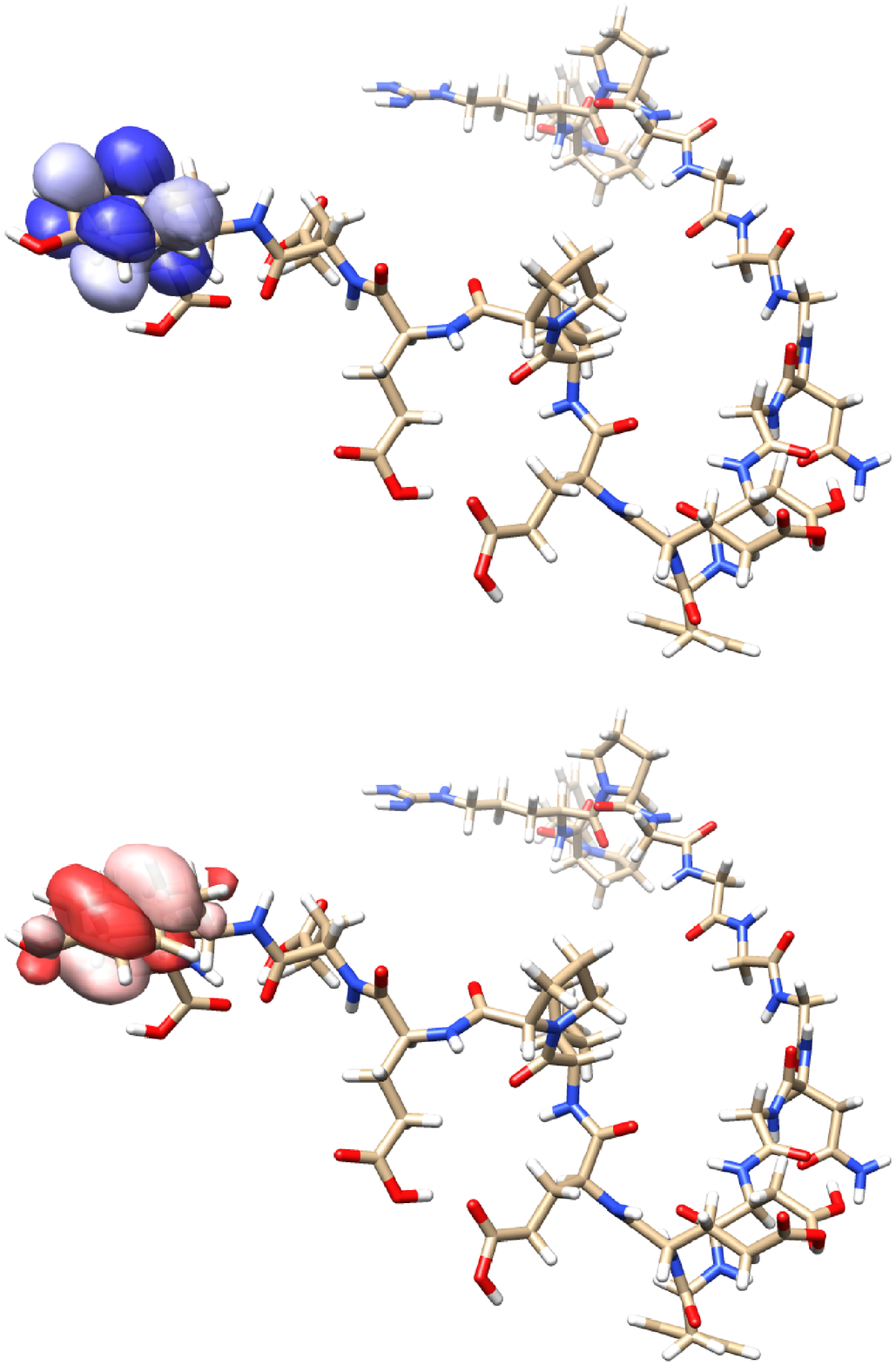}
	\caption{Stick representation of the bivalirudin molecule. Natural transition orbitals 
      for the lowest transition are represented with a contour value of $0.02$ a.u. 
      (bottom: occupied NTO, top: virtual NTO).\cite{chimera}
   }
   \label{fig:biva}
\end{figure}

In order to demonstrate the potential of \lofex{} for large molecules, we apply both the 
standard and spectrum strategies
for the calculation of the lowest excitation energy and the corresponding oscillator strength
of the bivalirudin molecule (see \Cref{fig:biva}). Bivalirudin is a synthetic polypeptide
containing $20$ residues. The structure used in this paper was obtained from the ChemSpider
database,\footnote{ChemSpider page for bivalirudin: CSID:10482069, \url{http://www.chemspider.com/Chemical-Structure.10482069.html} 
(accessed 11:56, Nov 21, 2016)} hydrogen atoms were added and the geometry
was relaxed at the molecular mechanics level (MMFF94\cite{Halgren1996} force field) using Avogadro.\cite{avogadro}\up{,}\footnote{Avogadro: 
an open-source molecular builder and visualization tool. Version 1.1.1. 
\url{http://avogadro.openmolecules.net/}} The Cartesian
coordinates of the optimized structure are available in the supporting information.\footnote{See 
supporting information at [URL will be inserted by AIP] for the molecular geometry of bivalirudin}
The calculations have been performed using the cc-pVDZ and aug-cc-pVDZ' basis sets
which (for the whole molecule) contain $2860$ and $4255$ basis functions, respectively.

\begin{table*}
   \caption{
      \lofex{} CC2 calculations of excitation energies and oscillator strengths for the lowest
      transition of the bivalirudin molecule. For standard-\lofex{}, $\tau_\omega = 0.02$ eV
      and we report values for $\omega^{(n)}$ and $f^{(n-1)}$, while for spectrum-\lofex{}, 
      $\tau_f = 0.001$ and  we report values for $\omega^{(n)}$ and $f^{(n)}$.
      For comparison TDDFT/CAM-B3LYP results are also reported.
      Timings are given in hours for all calculations and the fraction of time
      spent in the CC part of the \lofex{} calculations is given in \% as $T_\text{CC/tot}$. 
      \label{tab:biva}
   }
   \begin{ruledtabular}
   \begin{tabular}{llccccc}
      Basis set & method & No. iter. $(n)$ & $\omega$ & $f$ & Time (hours) & $T_\text{CC/tot}$ \\
      \hline
      \multirow{3}{*}{cc-pVDZ}      & standard-\lofex{} & 3 & 4.98 & 0.030 & 7  & 3.2 \\
                                    & spectrum-\lofex{} & 4 & 4.98 & 0.029 & 8  & 15 \\
                                    & CAM-B3LYP       & --- & 5.14 & 0.034 & 13 & ---  \\
      \hline
      \multirow{3}{*}{aug-cc-pVDZ'} & standard-\lofex{} & 3 & 4.82 & 0.028 & 157\footnote{
         For the aug-cc-pVDZ' results, the TDDFT calculation and the TDHF parts of the \lofex{} calculations
         were performed in parallel using 6 compute-nodes. Timings for those parts was therefore
         scaled by the number of nodes.
      }                                                              & 1.4 \\
        & spectrum-\lofex{} & 4 & 4.82 & 0.026 & 164\footnotemark[1] & 5.5 \\
        & CAM-B3LYP       & --- & 5.01 & 0.029 & 205\footnotemark[1] & ---  \\
   \end{tabular}
   \end{ruledtabular}
\end{table*}

One of the goals of \lofex{} is to provide CC results with a computational cost that can compete 
with TDDFT. In order to evaluate this feature for the bivalirudin calculations, we have performed
TDDFT/CAM-B3LYP\cite{Runge1984,Casida1995,Yanai2004} calculations using the same basis sets and targeting the same transition as for
the \lofex{} calculations. We note that for a fair comparison, the density-fitting\cite{Whitten1973,Baerends1973,Reine2008} approximation
for the Coulomb integrals was used in both the TDDFT calculations and in the TDHF part of \lofex{}.
We have also performed the \lofex{} calculations without density-fitting in the TDHF part and
verified that the final \lofex{}-CC2 results were not affected by this approximation (to the desired precision).
Note, that the calculations in \Cref{sec:res1} were performed without using density-fitting in the TDHF part of \lofex{}.
In \Cref{tab:biva}, we report timings for \lofex{} as well as for the TDDFT calculations.
For \lofex{}, we also report the fraction of the time (in \%) spent in the CC part of the calculations denoted 
$T_\text{CC/tot}$.
All the calculations reported in \Cref{tab:biva} were performed on Dell C6220 II compute-nodes, 
with 2 ten-core Intel E5-2680 v2 CPUs @ 2.8 GHz and 128 GB of memory.

Regarding the computational efforts in \lofex{}, the values for $T_\text{CC/tot}$ in \Cref{tab:biva} indicates that
only a few percents of the time is spent in the CC2 part of the calculations. In the best
case, for the standard-\lofex{}/aug-cc-pVDZ' result only $1.4$ \% is spent in the CC2 algorithm,
while $15$ \% are used in the spectrum-\lofex{}/cc-pVDZ calculation. Of course, for a given type
of transition, the larger the molecule, the smaller $T_\text{CC/tot}$ would be. 
This indicates that, as expected, \lofex{} effectively enables CC calculations of excitation energies
and oscillator strengths at roughly the cost of a TDHF calculation, provided that the transition
of interest is local compared to the size of the molecule.
In fact, the \lofex{} calculations are between $1.2$ and $1.9$ times faster than the corresponding TDDFT/CAM-B3LYP calculations.

In \Cref{tab:biva}, we also report the excitation energies and oscillator strengths obtained with
the different methods (TDDFT and \lofex{}). 
Both \lofex{} strategies give the same excitation energies for which 
a red-shift of $0.16$ eV is observed when adding diffuse functions in the basis set. 
The TDDFT numbers lie $0.16$ and $0.19$ eV higher than the CC2 excitation energies for the cc-pVDZ
and aug-cc-pVDZ' basis sets, respectively, which shows reasonably good agreement between the two methods.
As expected, the values for the oscillator strengths are slightly more dependent on the choice of the 
\lofex{} strategy. Since CC2 reference numbers are out of reach, one should consider the results of 
the spectrum-\lofex{} strategy to be superior (it takes one more iteration to converge). 
The TDDFT oscillator strengths are slightly higher for both basis sets but still very close to the CC2 results.


\section{Conclusion}

In this paper we have presented an extension of the \lofex{} algorithm to the computation of
oscillator strengths using CC2 linear-response theory. 
In \lofex{}, a state-specific mixed orbital space is generated from a TDHF calculation on the 
whole molecule by considering the dominant pair of NTOs, while the remaining orbitals are localized.
A reduced excitation orbital space (XOS), is then determined in a black-box manner for each electronic transition.
Two different strategies have been suggested for the computation of oscillator strengths within
\lofex{}: a standard strategy in which the XOS is optimized solely based on the CC2 excitation energy,
while the oscillator strength is only calculated in the converged (penultimate) XOS,
and a spectrum strategy which performs the XOS optimization directly based on the oscillator strength.
The first approach is designed to provide accurate excitation energies for all targeted transitions,
while the second strategy is dedicated to the calculation of electronic spectra, such that
strong transitions are described accurately, while less computational efforts are spent on weak 
and forbidden transitions.

Both strategies have shown promising results in terms of accuracy when applied to a set of 
medium-sized organic molecules. Significant computational savings with respect to conventional 
CC2 implementations are obtained whenever the considered transitions are local compared 
to the size of the molecule.
However, we note that for the strongest transition investigated in this work (S$_1$ of 11-cis-retinal),
no computational savings could be obtained due to the delocalized electronic structure of the molecule.
Many spectroscopically interesting chromophores have a delocalized electronic structure,\cite{Jacquemin2015}
and for such species, little or no computational savings would be obtained using \lofex{}.
In order to extend the applicability of \lofex{}, it might therefore be necessary
to further reduce the size of the XOS by considering, \eg{}, pair natural orbitals 
(PNOs),\cite{Helmich2013,Dutta2016} or improved NTOs.
This issue will be addressed in future publications.
Nonetheless, the current \lofex{} algorithm could be applied successfully to the bivalirudin molecule
with $4255$ basis functions, demonstrating that for transitions that are local compared to 
the size of the molecule, \lofex{} can provide CC2 excitation energies and oscillator strengths
at a computational cost competing with that of TDDFT.

\section*{Acknowledgments}

The research leading to these results has received funding from the European Research Council under the 
European Unions Seventh Framework Programme (FP/2007-2013)/ERC Grant Agreement no. 291371.

The numerical results presented in this work were performed at the Centre for Scientific Computing, Aarhus (\url{http://phys.au.dk/forskning/cscaa/}). 

This research used resources of the Oak Ridge Leadership Computing Facility at the Oak Ridge National Laboratory, which is supported by the Office of Science of the U.S. Department of Energy under Contract No. DE-AC05-00OR22725.

\appendix
\section{Working equations for CC2 transition moments}
\label{ap:cc2eq}

In this appendix we summarize the working equations of the CC2 model for the
calculation of excitation energies and (ground-state to excited-state) transition moments for closed-shell molecules.
For the derivation of those equations we have considered spin-free canonical orbitals and the 
following biorthonormal basis,\cite{mest}
\begin{subequations}
   \begin{align}
      \Bra{^{ab}_{ij}} &= \Bra{\text{HF}} E_{jb} E_{ia}, \\
      \Ket{^{ab}_{ij}} &= E_{ai} E_{bj} \Ket{\text{HF}}, \\
      \bibrad{a}{i}{b}{j} &= \frac{1}{1+\delta_{ai,bj}} \left( \frac{1}{3} \Bra{^{ab}_{ij}} + \frac{1}{6} \Bra{^{ab}_{ji}} \right), \\
      \Braket{\widetilde{^{ab}_{ij}} | ^{cd}_{kl}} &= \delta_{aibj,ckdl},
   \end{align}
\end{subequations}
where $E_{ai}$ is a singlet excitation operator in second-quantization.
The singles and doubles cluster operators are then defined as follows,
\begin{equation}
T_1 = \sum_{ai} t^{a}_{i} E_{ai},
\end{equation}
\begin{equation}
\label{eq:doubles_op}
T_2 = \frac{1}{2} \sum_{aibj} t^{ab}_{ij} E_{ai} E_{bj}.
\end{equation}

In the following section we only provide the CC2 working equations, for the details regarding the algorithm and the use of the RI approximation for the two-electron repulsion integrals, we refer to Refs. \onlinecite{Hattig2000,Hattig2002}.

\subsection{Overview}

The computation of transition moments from CC2 linear-response theory can be performed as follows 
(all the intermediate quantities are given in the following sections),

\begin{enumerate}
   \item Determine the ground-state singles amplitudes $t^a_i$ from Eq. (\ref{eq:amp1}) and using \Cref{tab:work1} (left).
   \item Determine the ground-state singles Lagrangian multipliers $\bar{t}^a_i$ from \Cref{eq:tbar} and using \Cref{tab:work2} (left) with $\omega = 0$ and
      with the right-hand-side from \Cref{eq:tbar_rhs}.
   \item Determine the ``right'' singles excitation amplitudes $R^a_i$ from \Cref{eq:xright} and using \Cref{tab:work1} (right).
   \item Determine the ``left'' singles excitation amplitudes $L^a_i$ from \Cref{eq:xleft} and using \Cref{tab:work2} (left).
   \item Check that right and left excitation energies agree to the desired precision 
      and normalize the excitation vectors using \Cref{eq:r2,eq:l2},
\begin{equation}
   \label{eq:norm}
\sum_{ai} L^a_i R^a_i + \frac{1}{2} \sum_{aibj} L^{ab}_{ij} R^{ab}_{ij} = 1
\end{equation}
   \item Determine the transition moment Lagrangian multipliers $\bar{M}^a_i$ from \Cref{eq:mbar} and using \Cref{tab:work2} (left). The right-hand-side
      has to be computed beforehand from \Cref{tab:work2} (right) which requires the optimized right excitation amplitudes
      and the ground-state Lagrangian multipliers. The corresponding doubles quantities are computed on-the-fly from \Cref{eq:r2,eq:r2sym,eq:tbar2}.
   \item Compute the one-particle density matrices given in \Cref{sec:dens} using the doubles quantities in \Cref{sec:doubles}.
   \item The density matrices can then be contracted with electric dipole moment T$_1$-transformed integrals to get the transition strengths as in \Cref{eq:trans,eq:rmom,eq:lmom}.
\end{enumerate}

\subsection{Integrals and Fock matrices}

We write two-electron repulsion integrals in the Mulliken notation as,
\begin{equation}
(pq|rs) = \sum_{\alpha \beta \gamma \delta} C_{\alpha p} C_{\beta q} C_{\gamma r} C_{\delta s} (\alpha \beta | \gamma \delta)
\end{equation}
where the $C_{\alpha p}$ are Hartree-Fock canonical MO coefficients.

A general inactive Fock matrix is given by 
\begin{align}
F_{pq} =& h_{pq} + \sum_{i} [2 (pq|ii) - (pi|iq)] = \delta_{pq} \epsilon_p \\
h_{pq} =& \sum_{\alpha \beta} C_{\alpha p} C_{\beta q} h_{\alpha \beta}
\end{align}
where we have introduced the one-electron integrals $h_{pq}$ and Hartree-Fock orbital energies $\epsilon_p, \epsilon_q\dots$

We consider integrals transformed with the singles ground-state amplitudes,
\begin{align}
(pq\hat{|}rs) =& \sum_{\alpha \beta \gamma \delta} X_{\alpha p} Y_{\beta q} X_{\gamma r} Y_{\delta s} (\alpha \beta | \gamma \delta) \\
\hat{h}_{pq} =& \sum_{\alpha \beta} X_{\alpha p} Y_{\beta q} h_{\alpha \beta}
\end{align}

\begin{equation}
   \begin{array}{rcl}
      X_{\alpha i} &=& C_{\alpha i} \\ 
      Y_{\alpha i} &=& C_{\alpha i} + \sum_a C_{\alpha a} t^a_i 
   \end{array}\qquad
   \begin{array}{rcl}
      X_{\alpha a} &=& C_{\alpha a} - \sum_i C_{\alpha i} t^a_i \\
      Y_{\alpha a} &=& C_{\alpha a}
   \end{array}
\end{equation}

We also have integrals transformed with a general ``right'' singles vector, $b^a_i$,
\begin{align}
(pq\bar{|}rs) =& P^{pr}_{qs} \sum_{\alpha \beta \gamma \delta}  
( \bar{X}_{\alpha p} Y_{\beta q} + X_{\alpha p} \bar{Y}_{\beta q} )X_{\gamma r} Y_{\delta s} (\alpha \beta | \gamma \delta) \\
P^{pr}_{qs} f^{pr}_{qs}  =& f^{pr}_{qs} + f^{rp}_{sq} \\
\bar{h}_{pq} =& \sum_{\alpha \beta} ( \bar{X}_{\alpha p} Y_{\beta q} + X_{\alpha p} \bar{Y}_{\beta q} ) h_{\alpha \beta}
\end{align}

\begin{equation}
\begin{array}{rcl}
\bar{X}_{\alpha i} &=& 0 \\
\bar{Y}_{\alpha i} &=& \sum_a C_{\alpha a} b^a_i 
\end{array}\qquad
\begin{array}{rcl}
\bar{X}_{\alpha a} &=& - \sum_i C_{\alpha i} b^a_i \\
\bar{Y}_{\alpha a} &=& 0 
\end{array}
\end{equation}
where, depending on the context, $b^a_i$ may correspond to the trial right excitation amplitudes 
or the optimized right excitation amplitudes $R^a_i$.

Similarly, we consider integrals transformed with a general ``left'' singles vector, $\bar{b}^a_i$,
\begin{align}
(pq\breve{|}rs) =& P^{pr}_{qs} \sum_{\alpha \beta \gamma \delta}  
( \breve{X}_{\alpha p} Y_{\beta q} + X_{\alpha p} \breve{Y}_{\beta q} )X_{\gamma r} Y_{\delta s} (\alpha \beta | \gamma \delta)
\end{align}

\begin{equation}
\begin{array}{rcl}
\breve{X}_{\alpha i} &=& \sum_a X_{\alpha a} \bar{b}^a_i \\
\breve{Y}_{\alpha i} &=& 0
\end{array}\qquad
\begin{array}{rcl}
\breve{X}_{\alpha a} &=& 0 \\
\breve{Y}_{\alpha a} &=& - \sum_i Y_{\alpha i} \bar{b}^a_i
\end{array}
\end{equation}
where, depending on the context, $\bar{b}^a_i$ may correspond to the trial left excitation amplitudes,
the optimized left excitation amplitudes $L^a_i$, 
the ground-state Lagrangian multipliers $\bar{t}^a_i$,
or the transition moment Lagrangian multipliers $\bar{M}^a_i$.

Finally, we also introduce the following one-index transformed integrals, 
\begin{align}
   \widetilde{(ia|jb)} = -P^{ab}_{ij} \left( \sum_{ck} \bar{t}^c_i R^c_k (ka|jb) + \sum_{ck} \bar{t}^a_k R^c_k (ic|jb) \right).
\end{align}

Expressions for the different blocks of the $T_1$-transformed and ``right''-transformed Fock matrices 
are given in \Cref{tab:fock}.

\begin{table*}
   \caption{Inactive transformed Fock matrices.}
   \label{tab:fock}
   \begin{ruledtabular}
      \begin{tabular}{ll}
         $\hat{F}_{pq} = \hat{h}_{pq} + \sum_i [2 (pq\hat{|}ii) - (pi\hat{|}iq)]$ 
         & $\bar{F}_{pq} = \bar{h}_{pq} + \sum_i [2 (pq\bar{|}ii) - (pi\bar{|}iq)]$ \\
         $\hat{F}_{ij} = \sum_{ck} [2 (ij\hat{|}kc) - (ic\hat{|}kj)] t^c_k + \epsilon_i \delta_{ij}$ 
         & $\bar{F}_{ij} = \sum_{kc} [2 (ij\hat{|}kc) - (ic\hat{|}kj)] R^c_k + \sum_b R^b_j \hat{F}_{ib}$ \\
         $\hat{F}_{ia} = \sum_{ck} [2 (ia|kc) - (ic|ka)] t^c_k $ 
         & $\bar{F}_{ia} = \sum_{kc} [2 (ia|kc) - (ic|ka)] R^c_k$ \\
         $\hat{F}_{ai} = \sum_{ck} [2 (ai\hat{|}kc) - (ac\hat{|}ki)] t^c_k + (\epsilon_a - \epsilon_i) t^a_i $
         & $\bar{F}_{ai} = \sum_{kc} [2 (ai\hat{|}kc) - (ac\hat{|}ki)] R^c_k + \sum_b R^b_j \hat{F}_{ab} - \sum_j R^a_j \hat{F}_{ji}$ \\
         $\hat{F}_{ab} = \sum_{ck} [2 (ab\hat{|}kc) - (ac\hat{|}kb)] t^c_k + \epsilon_a \delta_{ab}$ 
         & $\bar{F}_{ab} = \sum_{kc} [2 (ab\hat{|}kc) - (ac\hat{|}kb)] R^c_k - \sum_j R^a_j \hat{F}_{jb} $
      \end{tabular}
   \end{ruledtabular}
\end{table*}

\subsection{Linear-transformed vectors and right-hand-sides}

In \Cref{tab:work1} we gather the working equations for the ground-state singles residual,
\begin{equation}
\Omega_{ai} = \Omega_{ai}^0 + \Omega_{ai}^G + \Omega_{ai}^H + \Omega_{ai}^I + \Omega_{ai}^J  = 0,
\end{equation}
and for a ``right'' linear-transformed vector,
\begin{equation}
\sigma_{ai} = \sum_{bj} \Aeff_{ai,bj}(\omega) b^b_j 
			=  \sigma_{ai}^0 + \sigma_{ai}^G + \sigma_{ai}^H + \sigma_{ai}^I + \sigma_{ai}^J,
\end{equation}
while \Cref{tab:work2} contains the working equations for a ``left'' linear-transformed vector,
\begin{equation}
\bar{\sigma}_{ai} = \sum_{bj} \bar{b}^b_j \Aeff_{bj,ai}(\omega)  
			=  \bar{\sigma}_{ai}^0 + \bar{\sigma}_{ai}^G + \bar{\sigma}_{ai}^H 
            + \bar{\sigma}_{ai}^I + \bar{\sigma}_{ai}^J,
\end{equation}
and for the effective right-hand-side of the transition moment Lagrangian multipliers,
\begin{equation}
\bar{m}^\text{eff}_{ai} = \bar{m}^\text{eff,0}_{ai} + \bar{m}^\text{eff,G}_{ai} 
	+ \bar{m}^\text{eff,H}_{ai} + \bar{m}^\text{eff,I}_{ai} + \bar{m}^\text{eff,J}_{ai}.
\end{equation}

\begin{table*}
   \caption{
      CC2 working equations for the ground-state residual $\Omega_{ai}$ 
      and a ``right'' linear-transformed vector $\sigma_{ai}$.
   } 
   \label{tab:work1}
   \begin{ruledtabular}
      \begin{tabular}{lll}
         Terms & $\Omega_{ai} $ & $\sigma_{ai} = \Aeff_{ai,bj}(\omega) b^b_j$  \\
         \hline 
         $0$ & $(\epsilon_a - \epsilon_i) t^a_i$
         & $\sum_{b} E_{ab} b^b_i - \sum_{j} E_{ji} b^a_j$ \\
         $G$ & $+ \sum_{cdk} \tilde{t}^{dc}_{ik} (kc\hat{|}ad)$ 
         & $+ \sum_{cdk} \tilde{b}^{dc}_{ik} (kc\hat{|}ad)$ \\
         $H$ & $- \sum_{ckl} \tilde{t}^{ac}_{lk} (kc\hat{|}li)$ 
         & $- \sum_{ckl} \tilde{b}^{ac}_{lk} (kc\hat{|}li)$ \\
         $I$ & $+ \sum_{ck} \tilde{t}^{ac}_{ik} \hat{F}_{kc}$ 
         & $+ \sum_{ck} \left[ \tilde{b}^{ac}_{ik} \hat{F}_{kc} + \tilde{t}^{ac}_{ik} \bar{F}_{kc} \right]$ \\
         $J$ & $+ \sum_{ck} [2(kc\hat{|}ai) - (ki\hat{|}ac)] t^c_k$ 
         & $+ \sum_{ck} [2(kc\hat{|}ai) - (ki\hat{|}ac)] b^c_k$ \\
         \hline \\
         & $\tilde{t}^{ab}_{ij} = \frac{2 (ai\hat{|}bj) - (bi\hat{|}aj)}{\den{}}$ 
         & $\tilde{b}^{ab}_{ij} = \frac{2 (ai\bar{|}bj) - (bi\bar{|}aj)}{\den{} + \omega}$ \\
         & & $E_{ji} = \hat{F}_{ji} + \sum_{cdk} \tilde{t}^{dc}_{ik} (kc|jd)$ \\
         & & $E_{ab} = \hat{F}_{ab} - \sum_{ckl} \tilde{t}^{ac}_{lk} (kc|lb)$ \\
      \end{tabular}
   \end{ruledtabular}
\end{table*}

\begin{table*}
   \caption{
      CC2 working equations for a ``left'' linear-transformed vector $\bar{\sigma}_{ai}$ 
      and the right-hand-side of the transition moment Lagrangian multipliers equation $\bar{m}_{ai}^\text{eff}$.
   } 
   \label{tab:work2}
   \begin{ruledtabular}
      \begin{tabular}{lll}
         Terms & $\bar{\sigma}_{ai} = \bar{b}^b_j \Aeff_{bj,ai}(\omega)$ & $\bar{m}_{ai}^\text{eff}$ \\
         \hline 
         $0$ & $\sum_{b} E_{ba} \bar{b}^b_i - \sum_{j} E_{ij} \bar{b}^a_j$ 
         & $\sum_{b} \bar{E}_{ba} \bar{t}^b_i - \sum_{j} \bar{E}_{ij} \bar{t}^a_j$  \\
         $G$ & $+ \sum_{cdk} \bar{b}^{dc}_{ik} (ck\hat{|}da)$ 
         & $+ \sum_{cdk} [F^{dc}_{ik} (ck\hat{|}da) + \bar{t}^{dc}_{ik} (ck\bar{|}da)]$ \\
         $H$ & $- \sum_{ckl} \bar{b}^{ac}_{lk} (ck\hat{|}il)$ 
         & $- \sum_{ckl} [ F^{ac}_{lk} (ck\hat{|}il) + \bar{t}^{ac}_{lk} (ck\bar{|}il)]$ \\
         $I$ & $+ \sum_{ck} [2(kc|ia) - (ka|ic)] C^c_k$
         & $+ \sum_{ck} [2(kc|ia) - (ka|ic)] \bar{C}^c_k$\\
         & & $+ 2\sum_{ck} [ 2 (kc|ia) - (ka|ic) ] R^c_k$ \\
         $J$ & $+ \sum_{ck} [2(ck\hat{|}ia) - (ca\hat{|}ik)] \bar{b}^c_k$ 
         & $+ \sum_{ck} [2(ck\bar{|}ia) - (ca\bar{|}ik) ] \bar{t}^c_k$ \\
         \hline \\
         & $\bar{b}^{ab}_{ij} = \frac{2 (ia\breve{|}jb) - (ib\breve{|}ja) + P^{ab}_{ij} [ 2 \bar{b}^a_i \hat{F}_{jb} - \bar{b}^a_j \hat{F}_{ib} ]}
         {\den{} + \omega}$ 
         & $F^{ab}_{ij} = \frac{2 \widetilde{(ia|jb)} - \widetilde{(ib|ja)} + P^{ab}_{ij} [ 2 \bar{t}^a_i \bar{F}_{jb} - \bar{t}^a_j \bar{F}_{ib} ]}{\den{} - \omega}$ \\
         & $C^a_i = \sum_{bj} \tilde{t}^{ab}_{ij} \bar{b}^b_j$
         & $\bar{C}^a_i = \sum_{bj} \tilde{R}^{ab}_{ij} \bar{t}^b_j$ \\
           & $E_{ij} = \hat{F}_{ij} + \sum_{cdk} \tilde{t}^{dc}_{jk} (kc|id)$ 
         & $\bar{E}_{ij} = \bar{F}_{ij} + \sum_{cdk} \tilde{R}^{dc}_{jk} (kc|id)$ \\
           & $E_{ba} = \hat{F}_{ba} - \sum_{ckl} \tilde{t}^{bc}_{lk} (kc|la)$ 
         & $\bar{E}_{ba} = \bar{F}_{ba} - \sum_{ckl} \tilde{R}^{bc}_{lk} (kc|la)$ \\
      \end{tabular}
   \end{ruledtabular}
\end{table*}

The effective right-hand-side for the ground-state Lagrangian multipliers is given by,
\begin{align}
   \label{eq:tbar_rhs}
\eta^\text{eff}_{ai} =& 2 \hat{F}_{ia} + \sum_{ckd} \tilde{\eta}^{cd}_{ki}  (ck\hat{|}da) - \sum_{ckl} \tilde{\eta}^{ca}_{kl}  (ck\hat{|}il),
\end{align}

\begin{align}
   \label{eq:tbar_rhs2}
\tilde{\eta}^{ab}_{ij} = 2 \frac{2(ia|jb) - (ib|ja)}{\den{}}.
\end{align}

\subsection{Doubles quantities}
\label{sec:doubles}

All doubles quantities can be calculated on-the-fly from the corresponding singles which are kept in memory.
We consider the ground-state doubles amplitudes,
\begin{align}
t^{ab}_{ij} =& \frac{ (ai\hat{|}bj) }{\den{}},
\end{align}
the right doubles excitation amplitudes,
\begin{align}
   \label{eq:r2}
R^{ab}_{ij} =& \frac{ (ai\bar{|}bj) }{\den{} + \omega},
\end{align}
the left doubles excitation amplitudes,
\begin{align}
   \label{eq:l2}
L^{ab}_{ij} =& \frac{2 (ia\breve{|}jb) - (ib\breve{|}ja) + P^{ab}_{ij} [ 2 L^a_i \hat{F}_{jb} - L^a_j \hat{F}_{ib} ]}{\den{} + \omega},
\end{align}
the ground-state doubles Lagrangian multipliers,
\begin{align}
   \label{eq:tbar2}
\bar{t}^{ab}_{ij} =& \tilde{\eta}^{ab}_{ij} + \frac{2 (ia\breve{|}jb) - (ib\breve{|}ja) + P^{ab}_{ij} [ 2 \bar{t}^a_i \hat{F}_{jb} - \bar{t}^a_j \hat{F}_{ib} ]}{\den{}},
\end{align}
and the transition moment doubles Lagrangian multipliers,
\begin{align}
\bar{M}^{ab}_{ij} =& F^{ab}_{ij} + \frac{2 (ia\breve{|}jb) - (ib\breve{|}ja) + P^{ab}_{ij} [ 2 \bar{M}^a_i \hat{F}_{jb} - \bar{M}^a_j \hat{F}_{ib} ]}{\den{} - \omega}.
\end{align}

Where $\tilde{\eta}^{ab}_{ij}$ and $F^{ab}_{ij}$ have been defined in \Cref{eq:tbar_rhs2} and \Cref{tab:work2} (right), respectively.
The ground-states doubles amplitudes and the doubles excitation amplitudes are often used in the form,
\begin{align}
\tilde{t}^{ab}_{ij} =& 2 t^{ab}_{ij} - t^{ba}_{ij}, \\
   \tilde{R}^{ab}_{ij} =& 2 R^{ab}_{ij} - R^{ba}_{ij}. \label{eq:r2sym}
\end{align}

\subsection{One-particle density matrices}
\label{sec:dens}

The CC2 transition moments are calculated from the following one-particle density matrices,
\begin{align}
D_{ij}^{\xi} (\mathbf{X}) &= -\sum_{abk} X_{jk}^{ab} t_{ik}^{ab} \\
D_{ia}^{\xi} (\mathbf{X}) &= \sum_{ck} X^c_k \tilde{t}_{ik}^{ac} \\
D_{ai}^{\xi} (\mathbf{X}) &= X^a_i \\
D_{ab}^{\xi} (\mathbf{X}) &= \sum_{ijc} X_{ij}^{ac} t_{ij}^{bc} 
\end{align}
where $\mathbf{X}$ denotes either the ``left'' excitation amplitudes $\mathbf{L}$ or
the transition moment Lagrangian multipliers $\bar{\mathbf{M}}$.

Finally, we have,
\begin{align}
D^{\eta}_{ij}(\mathbf{R}) &= - \sum_a \bar{t}^a_j R^a_i - \sum_{abk} \bar{t}_{jk}^{ab} R_{ik}^{ab} \\
D^{\eta}_{ia}(\mathbf{R}) &= 2 R^a_i + \sum_{ck} \bar{t}^c_k \tilde{R}_{ik}^{ac}
- \sum_b \bigg(
\sum_{kjc} \bar{t}_{kj}^{bc} t_{kj}^{ac}
\bigg) R^b_i
- \sum_j \bigg(
\sum_{cbk} \bar{t}_{jk}^{cb} t_{ik}^{cb}
\bigg) R^a_j \\
D^{\eta}_{ai}(\mathbf{R}) &= 0 \\
D^{\eta}_{ab}(\mathbf{R}) &= \sum_i \bar{t}_{ai} R^b_i + \sum_{ijc} \bar{t}_{ij}^{ac} R_{ij}^{bc}.
\end{align}

%

\end{document}